\documentclass[reprint, amsmath, amssymb, aps]{revtex4-1}
\usepackage{graphicx}
\usepackage{dcolumn}
\usepackage{bm}
\usepackage{mathrsfs}
\usepackage{hyperref}
\usepackage{epstopdf}
\usepackage{amsmath}
\usepackage{amssymb}
\usepackage{mathrsfs}
\usepackage{diagbox}

\begin{document}
\title{Emergent elastic waves in skyrmion crystals with finite frequencies at long wavelength limit}
\author{Yangfan Hu}
 \email[Corresponding author.]{huyf3@mail.sysu.edu.cn}
\affiliation{Sino-French Institute of Nuclear Engineering and Technology, Sun Yat-sen University, 519082, Zhuhai, China}

\begin{abstract}
   A fundamental fact in solids is that the frequencies of elastic waves vanish as the wave number approaches zero\cite{6}. Here we theoretically show that this fact is overturned when studying the lattice vibration of skyrmion crystals\cite{1,2,3} (SkX), i.e., periodic alignment of topologically nontrivial spin solitons called magnetic skyrmions\cite{4,5}. As emergent crystals, SkX possess collective excitations called ``emergent phonons'', which describe dynamics of SkX caused by lattice vibration (resembling acoustical branches of ordinary phonons) and in-lattice vibration (resembling optical branches of ordinary phonons). We find that lattice vibration and in-lattice vibration of the emergent phonons in SkX are coupled even at long wavelength limit, such that multiple types of ``emergent elastic waves'' (modes causing lattice vibration of SkX) with finite frequencies exist. This phenomenon, which originates from the Berry phase form of kinetic energy, is generally true for emergent crystalline states of spins. Our results show that the dynamics of magnetic emergent crystals are intrinsically different from that of ordinary crystals.
\end{abstract}

\maketitle

Magnetic skyrmions\cite{1,4,5} are particle-like spin textures stabilized by the noncollinear Dzyaloshinsky-Moriya interaction (DMI) in chiral magnets. Skyrmions appear as basic units in an emergent world such that they usually condensate into a crystalline state\cite{1,2,9} called SkX. Similar to ordinary crystals, the presence of SkX induce a novel type of collective excitations according to previous studies\cite{13,18,19,22,23,31}, which we call it ``emergent phonons''. The significance of this novel type of collective excitations is twofold: on one hand, similar to phonons in ordinary crystals\cite{11}, they determine the dynamical behaviors of the system, as well as some of the most fundamental macroscopic properties such as the elasticity\cite{20} and heat capacity. On the other hand, they are intrinsically collective behaviors of spins. This electromagnetic nature permits direct interaction with static or dynamic electromagnetic fields, inducing exotic properties such as high electric current mobility\cite{12,13}, elasticity under electric\cite{14} and magnetic fields\cite{15}, which are promising to possible microwave- and magnonics-related application\cite{16}.

Phonons in ordinary crystals are divided into two categories known as the acoustical branches and the optical branches\cite{6}. At long wavelength limit, the acoustical branches, corresponding to elastic waves, describe vibrations of lattice as a whole and have vanishing frequencies, while the optical branches describe relative vibrations of atoms inside a lattice and have finite frequencies. Such a categorization is subtly used in previous analytical studies of emergent phonons in SkX. Zang et. al\cite{13} consider the acoustical branches of emergent phonons in SkX at long wavelength limit, and obtained a  $\omega \propto {{k}^{2}}$ form of dispersion relation. In their study, SkX are regarded as monatomic crystals whose deformation at long wavelength is described merely by a continuous emergent displacement field, representing vibration of the lattices as a whole. On the other hand, SkX can be described by a Fourier series\cite{1,17}, and the vibration of the Fourier magnitudes, seemingly not inducing any lattice deformation, can also cause a change of the field patterns of SkX. Schwarze\cite{18} et. al study the vibration of these Fourier magnitudes, recovering the three distinct modes termed clockwise (CW), counterclockwise (CCW) and breathing in the few GHz range observed in $Cu_2OSeO_3$\cite{19}, $MnSi$ and $Fe_{0.8}Co_{0.2}Si$\cite{18}. Nevertheless, the displacement of the lattice as a whole is not considered in their formulation, for which their analysis seems to be focusing on the optical branches of emergent phonons in SkX. A key issue unsettled in these studies is that at long wavelength limit, do the acoustical branches and the optical branches of the emergent phonons in SkX couple, or do they behave like independent vibrational modes such as those of ordinary crystals. By systematically studying the two types of vibrations together in SkX, we find that they are strongly coupled even at long wavelength limit, such that multiple modes exist which cause vibration of the lattice as a whole, while the so-called ``Goldstone mode'' cause simultaneously a rigid translation of the lattice and a distortion of the field pattern of skyrmion inside the lattice.

We study Bloch-type SkX in B20 chiral magnets, where the following Landau-Ginzburg functional describes the rescaled free energy density of the system\cite{1}
\begin{equation}
\begin{aligned}
\tilde{\phi }\left( \mathbf{m} \right)=&\sum\limits_{i=1}^{3}{{{\left( \frac{\partial \mathbf{m}}{\partial {{r}_{i}}} \right)}^{2}}+2}\mathbf{m}\cdot \left( \nabla \times \mathbf{m} \right)-2\mathbf{b}\cdot \mathbf{m}
\\&+\tilde{T}{{\mathbf{m}}^{2}}+{{\mathbf{m}}^{4}},
\label{1}
\end{aligned}
\end{equation}
where $\mathbf{m}$ is the rescaled magnetization, $\mathbf{b}$ is the rescaled magnetic field and $\tilde{T}$ is the rescaled temperature. The rescaling process is described in the Methods section. In this work we assume that $\mathbf{b}={{\left[ \begin{matrix}
   0 & 0 & b  \\
\end{matrix} \right]}^{T}}$, so that eq. (\ref{1}) is isotropic in the $x-y$ plane. The effect of anisotropy is discussed in a subsequent work of ours\cite{33}. Deformable SkX with long range order permits the following Fourier expansion of $\mathbf{m}$\cite{17,20}
\begin{equation}
\begin{aligned}
\mathbf{m}=\sum\limits_{\mathbf{l}}{{{\mathbf{m}}_{{{\mathbf{q}}^e_{\mathbf{l}}}}}{{e}^{\text{i}{{\mathbf{q}}_{\mathbf{l}}}(\varepsilon _{ij}^{e},\ {{\omega }^{e}})\cdot \mathbf{r}}}},
\label{2}
\end{aligned}
\end{equation}
where ${{\mathbf{m}}_{{{\mathbf{q}}^e_{\mathbf{l}}}}}$ denotes the Fourier magnitudes, ${{\mathbf{q}}^e_{\mathbf{l}}}={{l}_{1}}{{\mathbf{q}}^e_{\mathbf{1}}}+{{l}_{2}}{{\mathbf{q}}^e_{2}}$ where ${{l}_{1}}$ and ${{l}_{2}}$ are integers and ${{\mathbf{q}}^e_{\mathbf{1}}}$ and ${{\mathbf{q}}^e_{2}}$ are the basic reciprocal vectors of SkX, which are deformable under external disturbance. The deformation of ${{\mathbf{q}}^e_{\mathbf{l}}}$ is described by the emergent elastic strains $\varepsilon _{ij}^{e}$ and the emergent rotational angle ${{\omega }^{e}}$\cite{20}, which are defined from the emergent displacement field ${{\mathbf{u}}^{e}}={{\left[ \begin{matrix}
   u_{1}^{e} & u_{2}^{e}  \\
\end{matrix} \right]}^{T}}$ by $\varepsilon _{ij}^{e}=\frac{1}{2}(u_{i,j}^{e}+u_{j,i}^{e})$ and ${{\omega }^{e}}=\frac{1}{2}(u_{1,2}^{e}-u_{2,1}^{e})$. Similar to atomic lattice, rigid translation of SkX does not induce a change of free energy, thus it is $\varepsilon _{ij}^{e}$ and ${{\omega }^{e}}$ instead of ${{\mathbf{u}}^{e}}$ which matter in the expression of $\mathbf{m}$. The expression of ${{\mathbf{q}}_{\mathbf{l}}}$ in terms of $\varepsilon _{ij}^{e}$ and ${{\omega }^{e}}$ depends on the crystalline structure of SkX, and is introduced in the Methods section for hexagonal SkX.

At appropriate condition of $b$ and $\tilde{T}$, minimization of the free energy based on eq. (\ref{1}) yields a metastable SkX\cite{5, 17}, while the thermodynamically stable state corresponds to the conical phase. Nevertheless, concerning the exotic robustness of metastable SkX observed in a wide area of the temperature-magnetic field phase diagram of helimagnets\cite{3,21}, the analysis of this metastable SkX phase is not only of theoretical interest but also with practical significance. 

From eq. (\ref{2}), we can describe the equilibrium magnetization of a metastable SkX by ${{\left( {{\mathbf{m}}_{{{\mathbf{q}}_{\mathbf{l}}}}} \right)}_{st}}$, ${{\left( \varepsilon _{ij}^{e} \right)}_{st}}$ and ${{\left( {{\omega }^{e}} \right)}_{st}}$. Consider a small vibration around this metastable state, which induces simultaneously a vibration of ${{\mathbf{u}}^{e}}$ denoted by ${{\left( {{\mathbf{u}}^{e}} \right)}_{v}}(\mathbf{r},t)$ and a vibration of ${{\mathbf{m}}_{{{\mathbf{q}}_{\mathbf{l}}}}}$ denoted by ${{\left( {{\mathbf{m}}_{{{\mathbf{q}}_{\mathbf{l}}}}} \right)}_{v}}(\mathbf{r},t)$. In this case, eq. (\ref{2}) becomes
\begin{equation}
\begin{aligned}
\mathbf{m}=\sum\limits_{\mathbf{l}}{\left[ {{\left( {{\mathbf{m}}_{{{\mathbf{q}}^e_{\mathbf{l}}}}} \right)}_{st}}+{{\left( {{\mathbf{m}}_{{{\mathbf{q}}_{\mathbf{l}}}}} \right)}_{v}} \right]{{e}^{\text{i}\left({\mathbf{q}}_{\mathbf{l}}\right)_{st}\cdot \left[ \mathbf{r}-{{\left( {{\mathbf{u}}^{e}} \right)}_{v}} \right]}}},
\label{3}
\end{aligned}
\end{equation}
where $\left({\mathbf{q}}^e_{\mathbf{l}}\right)_{st}={\mathbf{q}}^e_{\mathbf{l}}\left[ {{\left( \varepsilon _{ij}^{e} \right)}_{st}},\ {{\left( {{\omega }^{e}} \right)}_{st}} \right]$. It is convenient to write components of all Fourier magnitudes ${{\mathbf{m}}_{{{\mathbf{q}}_{\mathbf{l}}}}}$ in a single vector ${{\mathbf{m}}^{q}}$, for which the two vectors ${{\left( {{\mathbf{u}}^{e}} \right)}_{v}}$ and ${{\left( {{\mathbf{m}}^{q}} \right)}_{v}}$  include all the variables to be solved. In previous theoretical studies of the emergent elastic waves\cite{13,22}, ${{\left( {{\mathbf{u}}^{e}} \right)}_{v}}$ is considered solely, while in previous theoretical studies of the internal modes of SkX\cite{23}, ${{\left( {{\mathbf{m}}^{q}} \right)}_{v}}$ is considered solely. To establish an analytical framework which determines the collective spin excitations in SkX, we have to derive the coupled Euler-Lagrange equations for ${{\left( {{\mathbf{u}}^{e}} \right)}_{v}}$ and ${{\left( {{\mathbf{m}}^{q}} \right)}_{v}}$, which can be obtained from the least action principle by using eq. (\ref{3}) (See the Methods section for details). Consider the plane-wave form of solution ${{\left( {{\mathbf{u}}^{e}} \right)}_{v}}={{\mathbf{u}}^{e0}}{{e}^{\text{i}(\omega t-\mathbf{\tilde{k}}\cdot \mathbf{r})}}$, ${{\left( {{\mathbf{m}}^{q}} \right)}_{v}}={{\mathbf{m}}^{q0}}{{e}^{\text{i}(\omega t-\mathbf{\tilde{k}}\cdot \mathbf{r})}}$, a generalized eigenvalue problem of the frequency $\omega $ can be obtained at long wavelength limit
\begin{equation}
\begin{aligned}
\left( \mathbf{R}\omega -\mathbf{K} \right)\left[ \begin{matrix}
   {{\mathbf{u}}^{e0}}  \\
   {{\mathbf{m}}^{q0}}  \\
\end{matrix} \right]=\mathbf{0},
\label{4}
\end{aligned}
\end{equation}
where the expression of $\mathbf{R}$ and $\mathbf{K}$ are derived in the Methods section. Solution of eq. (\ref{4}) determines the dispersion relation of the coupled vibration of ${{\left( {{\mathbf{u}}^{e}} \right)}_{v}}$ and ${{\left( {{\mathbf{m}}^{q}} \right)}_{v}}$, as well as the eigenvectors for different modes. We denote the frequencies of different modes by ${{\omega }_{i}}(\mathbf{\tilde{k}})$ (${{\tilde{\omega }}_{i}}(\mathbf{\tilde{k}})={{\omega }_{i}}(\mathbf{\tilde{k}})/\eta $, where $\eta $ is a material dependent factor such that ${{\tilde{\omega }}_{i}}(\mathbf{\tilde{k}})$ is material independent), ordered in such a way that ${{\omega }_{1}}(\mathbf{0})<{{\omega }_{2}}(\mathbf{0})<{{\omega }_{3}}(\mathbf{0})<\cdots $.

We study the coupled vibration of ${{\left( {{\mathbf{u}}^{e}} \right)}_{v}}$ and ${{\left( {{\mathbf{m}}^{q}} \right)}_{v}}$ at the long wavelength limit. The real-space magnetization distribution undergoing the first 6 modes calculated at $\tilde{T}=0.5,\ b=0.3$, ${{\tilde{k}}_{1}}={{10}^{-\text{5}}},\ {{\tilde{k}}_{2}}=0$ is shown in FIG. 1, resembling those of a previous study\cite{24}, except that ${{\left( {{\mathbf{u}}^{e}} \right)}_{v}}$ is neglected in their work and considered here. The experimentally confirmed\cite{19,23} counter clockwise (CCW) mode (${{\omega }_{3}}$), clockwise (CW) mode (${{\omega }_{6}}$) and breathing mode (${{\omega }_{5}}$) all shows up, and the variation of their frequencies with the applied magnetic field plotted in FIG. 2 agree with corresponding experimental results\cite{24} (see Supplementary Videos 1-6 for the motion of these modes).

Unexpectedly, we find that even at the $\Gamma $ point, there is still strong coupling between ${{\left( {{\mathbf{u}}^{e}} \right)}_{v}}$ and ${{\left( {{\mathbf{m}}^{q}} \right)}_{v}}$, such that 3 out of the first 6 modes cause vibration of the lattices as a whole, while 2 of them possess nonzero frequency. When we consider more modes by incorporating a higher order Fourier expansion of $\mathbf{m}$ in eq. (\ref{3}), we find that 7 out of the first 20 modes have similar properties, which implies that a finite proportion of all modes present coupled vibration of ${{\left( {{\mathbf{u}}^{e}} \right)}_{v}}$ and ${{\left( {{\mathbf{m}}^{q}} \right)}_{v}}$ at the $\Gamma $ point. 
Meanwhile, we find that the ``Goldstone mode'' (${{\omega }_{\text{1}}}$), i.e., the emergent elastic wave with vanishing frequency at the $\Gamma$ point, has nonzero components of ${{{\mathbf{m}}^{q0}} }$ in its eigenvector of vibration. It means that the Goldstone mode (${{\omega }_{\text{1}}}$) no-longer describes pure rigid translational vibration of SkX, but is accompanied by deformation of the field pattern of skyrmion inside the lattice (as illustrated by Fig. 1(a1-a4)). This Goldstone mode solved by considering coupling between ${{\left( {{\mathbf{u}}^{e}} \right)}_{v}}$ and ${{\left( {{\mathbf{m}}^{q}} \right)}_{v}}$ still possesses a $\omega \propto {{k}^{2}}$ form of dispersion relation at long wavelength limit, yet with a smaller coefficient than the one obtained by solving the vibration of ${{\left( {{\mathbf{u}}^{e}} \right)}_{v}}$ alone (See the phonon spectrum plotted in FIG. 3 for comparison).  Actually, this accompanied “particle shape” deformation of the Goldstone mode has long been confirmed when studying the dynamics of isolated skyrmions\cite{25,26,29} and SkX\cite{30}. In isolated skyrmions the particle shape deformation of skyrmion in the Goldstone mode is found to induce a large “inertia” of the skymion and significantly affects its dynamics. Our results also explain why the experimentally obtained “inertia” of isolated skyrmion\cite{26} significantly exceeds that of the related theoretical prediction\cite{25}. Actually the vibrational modes studied in the two works are different: the experimentally excited modes\cite{26} in the GHz range corresponds to the CCW and CW modes of isolated skyrmion\cite{27}, while the theoretically studied one is the Goldstone mode\cite{25}. The magnitude of internal deformation in the Goldstone mode is much smaller than that of the CCW and CW modes, which dominates the effective mass of the “skyrmion particle” during motion and thus explains the discrepancy. Moreover, by analyzing its eigenvector of vibration, we find that this Goldstone mode no longer possesses two vibration direction of the lattices (corresponding to the longitudinal and transverse elastic waves of ordinary crystals), but has a unique mode of vibration where the longitudinal motion and transverse motion of lattices are coupled. In this case, the lattices of SkX undergo clockwise elliptical rotation around their equilibrium position (Figure 1(a1-a4) and Supplementary Video 1). The particle shape deformation and rotational motion of the Goldstone mode presented in Supplementary Video 1 are observed in simulation of the current-driven motion of SkX\cite{30}.
The physical origin of this exotic finite frequency emergent elastic waves in SkX lies in the berry phase form of the kinetic energy. To be more specific, it is caused by coupling terms such as $\frac{d}{dt}\left( u_{i}^{e}m_{i}^{c} \right),\ (i=1,\ 2)$ appear in the kinetic energy density, where $m_{i}^{c},\ (i=1,\ 2)$ denote components of the constant vector in the Fourier representation of rescaled magnetization. As a result, the motion of $u_{i}^{e},\ (i=1,\ 2)$ and $m_{i}^{c},\ (i=1,\ 2)$ are coupled in the mass matrix $\mathbf{R}$ in eq. (\ref{4}). Although in the stiffness matrix $\mathbf{K }$, the coupling between $u_{i}^{e},\ (i=1,\ 2)$ and ${{\left( {{\mathbf{m}}^{q}} \right)}_{v}}$ vanishes at the $\Gamma$ point, $m_{i}^{c},\ (i=1,\ 2)$ is coupled with half of the components in ${{\left( {{\mathbf{m}}^{q}} \right)}_{v}}$, which together render a strong coupling between ${{\left( {{\mathbf{u}}^{e}} \right)}_{v}}$ and ${{\left( {{\mathbf{m}}^{q}} \right)}_{v}}$. The above analysis shows that this feature should be common all emergent crystalline states composed of periodic spin textures.

Nonzero $u_{i}^{e0},\ (i=1,\ 2)$ components in the eigenvector of a vibrational mode means that when the mode is excited, global displacement of the lattice as a whole can be achieved. An immediate consequence of this exotic finite frequency emergent elastic waves is that AC external fields with different frequencies may be used to transport SkX by exciting different emergent elastic waves. Since these modes corresponds to different internal deformation pattern of SkX, we should expect different dynamic behaviors when the SkX is moved by different modes. Consider AC magnetic field as an example, an important question is which of all the calculated modes can be excited. This depends on whether the considered mode has a nonzero component of $m_{i}^{c},\ (i=1,\ 2,\ 3)$. E.g., CCW and CW modes are excited by an in-plane ac-magnetic field\cite{19,23} since they have nontrivial $m_{1}^{c0}$ and $m_{2}^{c0}$ in their eigenvector of vibration. The breathing mode is excited by an out-of-plane ac-magnetic field since they have nontrivial $m_{3}^{c0}$ in its eigenvector of vibration. In Table 1, we list the value of components of ${{\mathbf{u}}^{e0}}$ and ${{\mathbf{m}}^{c0}}$ in the unit eigenvector for the first 20 modes calculated at $\tilde{T}=0.5$, $b=0.3$ near the $\Gamma$ point. The variation of these components for the first 6 modes with the applied magnetic field is plotted in FIG. 2(b, c), which shows that the eigenvector of modes can be tuned by changing the bias magnetic field.

Our study shows that the emergent elastic wave excitations of emergent crystals in chiral magnets are fundamentally different from those of ordinary crystals, for which the incorporation of emergent lattice deformation and displacements when studying their collective spin excitations is generally indispensable. Further research in this direction not only helps us in understanding the discrepancies between emergent crystalline states and ordinary crystals in terms of their dynamics, but may also stimulate novel applications. We construct a reliable and analytical framework to study the collective spin excitations in magnetic emergent crystals, which can easily be extended to cases where they are deformed\cite{33}.

\section{Methods}
\
\\
$Free$ $energy$ $density$ $functional$ $of$ $B20$ $chiral$ $magnets$ $and$ $its$ $rescaling$\\
We use the following free energy density functional to study magnetic skyrmions in cubic helimagnets
\begin{equation}
\begin{aligned}
\phi \left( \mathbf{M} \right)=&\sum\limits_{i=1}^{3}{A{{\left( \frac{\partial \mathbf{M}}{\partial {{x}_{i}}} \right)}^{2}}+D}\mathbf{M}\cdot \left( \nabla \times \mathbf{M} \right)\\&-\mathbf{B}\cdot \mathbf{M}+\alpha (T-{{T}_{0}}){{\mathbf{M}}^{2}}+\beta {{\mathbf{M}}^{4}},
\label{5}
\end{aligned}
\end{equation}
where $\mathbf{M}$ denotes the magnetization, $\mathbf{B}$ denotes the magnetic field, and $T$ denotes the temperature. The terms on the rhs. of eq. (\ref{5}) denote respectively the exchange energy density with a coefficient $A$, the Dzyaloshinskii-Moriya interaction (DMI) with a coefficient $D$, the Zeeman energy density, and the second and fourth order Landau expansion terms. Eq. (\ref{5}) can be simplified by rescaling the spatial variables as $\mathbf{r}=\frac{\mathbf{x}}{{{L}_{D}}}$, $\mathbf{m}=\frac{\mathbf{M}}{{{M}_{0}}}$, $\mathbf{b}=\frac{\mathbf{B}}{{{B}_{0}}}$, $\tilde{T}=\frac{\alpha (T-{{T}_{0}})}{K}$, and ${{L}_{D}}=\frac{2A}{D}$, $K=\frac{{{D}^{2}}}{4A}$, ${{M}_{0}}=\sqrt{\frac{K}{\beta }}$, ${{B}_{0}}=2K{{M}_{0}}$, which yield $\phi \left( \mathbf{M} \right)=\frac{{{K}^{2}}}{\beta }\tilde{\phi }\left( \mathbf{m} \right)$, while $\tilde{\phi }\left( \mathbf{m} \right)$ is given in eq. (\ref{1}). The benefit of eq. (\ref{1}) compared with eq. (\ref{5}) is that it provides a free energy density functional that is independent of material parameters, so that results obtained from analyzing eq. (\ref{1}) has general significance to Bloch-type SkX in any B20 helimagnets.
\
\\
\\
$Fourier$ $representation$ $of$ $SkX$ $and$ $free$ $energy$ $minimization$\\
To determine a metastable SkX state at given $\tilde{T}$ and $b$, one has to substitute the analytical expression of $\mathbf{m}$ for the SkX phase into eq. (\ref{1}), and minimize the free energy of the system with respect to all independent variables. In practice, we use the following Fourier representation of SkX instead of eq. (\ref{2})\cite{20}
\begin{equation}
    \begin{aligned}
        \mathbf m=& \mathbf m^c+\sum^n_{i=1}\sum^{n_i}_{j=1}\mathbf m_{\mathbf q_{ij}} e^{{\rm i}[\mathbf I-\mathbf F^e(\mathbf r)]^T\mathbf q_{ij}\cdot\mathbf r},
    \end{aligned}
    \label{9a}
\end{equation}
where $\mathbf m^c$ denotes a constant vector, $F_{ij}^e(\mathbf r)=\varepsilon^e_{ij}+\omega^e_{ij}=u^e_{i,j}$, and ${{\mathbf{q}}_{ij}}$ denote the undeformed wave vectors organized according to the following rules: $\left| {{\mathbf{q}}_{1j}} \right|<\left| {{\mathbf{q}}_{2j}} \right|<\left| {{\mathbf{q}}_{3j}} \right|<......$, and $\left| {{\mathbf{q}}_{i1}} \right|=\left| {{\mathbf{q}}_{i2}} \right|=......=\left| {{\mathbf{q}}_{i{{n}_{i}}}} \right|$. When truncated at a specific value of $n$, the $n^{th}$ order Fourier representation given in eq. (\ref{9a}) saves all the significant Fourier terms up to the $n^{th}$ order, which is hard to achieve if one uses eq. (\ref{2}). 

It is convenient to expand $\mathbf m_{\mathbf q_{ij}}$ as $\mathbf m_{\mathbf q_{ij}}=c_{ij1}\mathbf P_{ij1}+c_{ij2}\mathbf P_{ij2}+c_{ij3}\mathbf P_{ij3}$, where $c_{ij1}=c^{re}_{ij1}+{\rm i}c^{im}_{ij1}$, $c_{ij2}=c^{re}_{ij2}+{\rm i}c^{im}_{ij2}$, $c_{ij3}=c^{re}_{ij3}+{\rm i}c^{im}_{ij3}$ are complex variables to be determined, and $\mathbf P_{ij1}=\frac{1}{\sqrt{2}s_iq}[-{\rm i}q_{ijy}, {\rm i}q_{ijx}, s_iq]^T$, $\mathbf P_{ij2}=\frac{1}{s_iq}[q_{ijx}, q_{ijy}, 0]^T$, $\mathbf P_{ij3}=\frac{1}{\sqrt{2}s_iq}[{\rm i}q_{ijy}, -{\rm i}q_{ijx}, s_iq]^T$ with $\mathbf q_{ij}=[q_{ijx}, q_{ijy}]^T$, $|\mathbf q_{ij}|=s_iq$. For 2-D hexagonal SkX, we can assume without loss of generality that $\mathbf q_{11}=[0,1]^T$, $\mathbf q_{12}=[-\frac{\sqrt{3}}{2},-\frac{1}{2}]^T$, and $q=1$. Comparing eq. (\ref{9a}) and eq. (\ref{2}), we have ${\mathbf{q}}^e_{ij}(\varepsilon _{ij}^{e},\ {{\omega }^{e}})=[\mathbf I-(\mathbf F^e(\mathbf r))^T]{\mathbf{q}}_{ij}$, which gives for the basic reciprocal vectors $\mathbf q^e_{11}=[-\varepsilon _{12}^{e}+{\omega }^{e},1-\varepsilon _{22}^{e}]^T$, $\mathbf q^e_{12}=[-\frac{\sqrt{3}}{2}(1-\varepsilon _{11}^{e})+\frac{1}{2}(\varepsilon _{12}^{e}-{\omega }^{e}),\frac{\sqrt{3}}{2}(\varepsilon _{12}^{e}+{\omega }^{e})-\frac{1}{2}(1-\varepsilon _{22}^{e})]^T$.

In this case, all the independent variables describing the rescaled magnetization of the SkX phase can be gathered in two vectors, which are given by
\begin{equation}
    \begin{aligned}
    {{\pmb{\varepsilon}}^{ea}}={{\left[ \varepsilon _{11}^{e},\ \varepsilon _{22}^{e},\ \varepsilon _{12}^{e},\ \omega^{e} \right]}^{T}},
    \label{9b}
    \end{aligned}
\end{equation}
and
\begin{equation}
\begin{aligned}
{{\mathbf{m}}^{q}}=[& {{m}^c_{1}},\ {{m}^c_{2}},\ {{m}^c_{3}},c_{111}^{re},c_{112}^{re},c_{113}^{re},c_{121}^{re},c_{122}^{re},\\
& c_{123}^{re},c_{131}^{re},c_{132}^{re},c_{133}^{re},c_{111}^{im},c_{112}^{im},c_{113}^{im}\cdots]^{T},
\label{10}
\end{aligned}
\end{equation}
where the length of $\mathbf{m}^{q}$ depends on the order of Fourier representation used. For $1^{st}$, $2^{nd}$, and $3^{rd}$ order Fourier representation, the length of $\mathbf{m}^{q}$ is 21, 39 and 57, respectively. At given $\tilde{T}$ and $b$, minimization of the free energy based on eq. (\ref{9a}) determines the equilibrium value of the two vectors ${{\pmb{\varepsilon}}^{ea}}$ and $\mathbf{m}^{q}$ for a metastable SkX phase, which are denoted by $({{\pmb{\varepsilon}}^{ea}})_{st}$ and $(\mathbf{m}^{q})_{st}$.
\
\\
\\
$Euler-Lagrange$ $equation$ $for$ $the$ $emergent$ $displacements$ $and$ $the$ $Fourier$ $magnitudes$ $of$ $SkX$\\
Here we briefly introduce the procedure to deduce the basic equations describing the coupled wave motion of ${{\left( {{\mathbf{u}}^{e}} \right)}_{v}}(\mathbf{r},t)$ and  ${{\left( {{\mathbf{m}}_{{{\mathbf{q}}_{\mathbf{l}}}}} \right)}_{v}}(\mathbf{r},t)$\cite{32}. The action of any cubic helimagnet undergoing small vibration of magnetization reads
\begin{equation}
\begin{aligned}
S={{S}_{K}}-\int{\Phi }dt
\label{6}
\end{aligned}
\end{equation}
where $\Phi=\int{\phi}dV$ denotes the free energy of the system, and ${{S}_{K}}=\int{{{E}_{K}}dt}$ derives from the Berry phase action of a spin\cite{28}, whose variational form reads
\begin{equation}
\begin{aligned}
\delta {{S}_{K}}=\frac{{{M}}}{\gamma }\int{\int{(\mathbf{n}\times \mathbf{\dot{n}})}\delta \mathbf{n}dVdt}.
\label{7}
\end{aligned}
\end{equation}
Here $\mathbf{n}$ denotes the unit vector of magnetization, ${{M}}$ the averaged modulus of magnetization and $\gamma $ the gyromagnetic ratio. It is convenient to replace $\mathbf{n}$ in eq. (\ref{7}) by the rescaled magnetization $\mathbf{m}$, which gives
\begin{equation}
\begin{aligned}
\delta {{S}_{K}}=\frac{{{M}}}{\gamma {{m}^{3}}}\int{\int{(\mathbf{m}\times \mathbf{\dot{m}})}\delta \mathbf{m}dVdt}.
\label{8}
\end{aligned}
\end{equation}
where the value of $m=\left| \mathbf{m} \right|$ depends on the rescaled magnetic field $\mathbf{b}$ and the rescaled temperature $\tilde{T}$. $\Phi =\int_{V}{\phi }dV$ denotes the free energy of the system, where $\phi \left( \mathbf{M} \right)=\frac{{{K}^{2}}}{\beta }\tilde{\phi }\left( \mathbf{m} \right)$ is deduced above and $\tilde{\phi }\left( \mathbf{m} \right)$ is given in eq. (\ref{1}). The rescaled magnetization for deformable SkX can be expanded as follow instead of eq. (\ref{3})
\begin{equation}
    \begin{aligned}
        \mathbf m=& (\mathbf m^c)_{st}+(\mathbf m^c)_{v}+\sum^\infty_{i=1}\sum^{n_i}_{j=1}[(\mathbf m_{\mathbf q_{ij}})_{st}+(\mathbf m_{\mathbf q_{ij}})_{v}]\\
        &\times e^{{\rm i}[\mathbf I-(\mathbf F^e(\mathbf r))_{st}]^T\mathbf q_{ij}\cdot[\mathbf r-(\mathbf u^e)_{v}]},
    \end{aligned}
    \label{9}
\end{equation}
where $(F_{ij}^e(\mathbf r))_{st}=(\varepsilon^e_{ij})_{st}+(\omega^e_{ij})_{st}$.

The Euler-Lagrange equations of ${{\left( {{\mathbf{u}}^{e}} \right)}_{v}}$ and ${{\left( {{\mathbf{m}}^{q}} \right)}_{v}}$ read
\begin{equation}
\begin{aligned}
-\frac{d}{dt}\left[ \frac{\partial \mathscr{L}}{\partial {{\left( {{{\mathbf{\dot{u}}}}^{e}} \right)}_{v}}} \right]+\frac{\partial \mathscr{L}}{\partial {{\left( {{\mathbf{u}}^{e}} \right)}_{v}}}-\sum\limits_{i}{\frac{d}{d{{r}_{i}}}\left[ \frac{\partial \mathscr{L}}{\partial {{\left( \mathbf{u}_{,i}^{e} \right)}_{v}}} \right]}=0,
\label{11}
\end{aligned}
\end{equation}
\begin{equation}
\begin{aligned}
-\frac{d}{dt}\left[ \frac{\partial \mathscr{L}}{\partial {{\left( {{{\mathbf{\dot{m}}}}^{q}} \right)}_{v}}} \right]+\frac{\partial \mathscr{L}}{\partial {{\left( {{\mathbf{m}}^{q}} \right)}_{v}}}-\sum\limits_{i}{\frac{d}{d{{r}_{i}}}\left[ \frac{\partial \mathscr{L}}{\partial {{\left( \mathbf{m}_{,i}^{q} \right)}_{v}}} \right]}=0,
\label{12}
\end{aligned}
\end{equation}
where $\mathscr{L}={{E}_{K}}-\Phi $ is the Lagrangian of the system, and $\mathbf{u}_{,i}^{e}=\frac{\partial {{\mathbf{u}}^{e}}}{\partial {{r}_{i}}}$. To actually use eqs. (\ref{11}, \ref{12}) for small vibration of ${{\left( {{\mathbf{u}}^{e}} \right)}_{v}}$ and ${{\left( {{\mathbf{m}}^{q}} \right)}_{v}}$, we first expand the averaged rescaled free energy density  $\bar{\phi }=\frac{1}{V}\int{\tilde{\phi }\left( \mathbf{m} \right)}dV$ in terms of ${{\left( {{\mathbf{u}}^{e}} \right)}_{v}}$ and ${{\left( {{\mathbf{m}}^{q}} \right)}_{v}}$ and their derivatives and retain the lowest order terms. Substituting ${{\left( {{\mathbf{u}}^{e}} \right)}_{v}}={{\mathbf{u}}^{e0}}{{e}^{\text{i}(\mathbf{\tilde{k}}\cdot \mathbf{r}-\omega t)}}$, ${{\left( {{\mathbf{m}}^{q}} \right)}_{v}}={{\mathbf{m}}^{q0}}{{e}^{\text{i}(\mathbf{\tilde{k}}\cdot \mathbf{r}-\omega t)}}$ into eqs. (\ref{11}, \ref{12}), eq. (\ref{4}) can be derived at long wavelength limit, with
\begin{equation}
\begin{aligned}
\mathbf{R}=\left[ \begin{matrix}
   {{\mathbf{R}}^{e}} & {{\mathbf{R}}^{eq}}  \\
   {{\left( {{\mathbf{R}}^{eq*}} \right)}^{T}} & {{\mathbf{R}}^{q}}  \\
\end{matrix} \right],
\label{13}
\end{aligned}
\end{equation}
\begin{equation}
\begin{aligned}
\mathbf{K }=\left[ \begin{matrix}
   {{\mathbf{K }}^{e}} & {{\mathbf{K }}^{eq}}  \\
   {{\left( {{\mathbf{K }}^{eq*}} \right)}^{T}} & {{\mathbf{K }}^{q}}  \\
\end{matrix} \right],
\label{14}
\end{aligned}
\end{equation}
where ${{\mathbf{R}}^{eq*}}$ and ${{\mathbf{K }}^{eq*}}$ denote complex conjugate of ${{\mathbf{R}}^{eq}}$ and ${{\mathbf{K }}^{eq}}$. In eqs. (\ref{13}, \ref{14}), $R_{ij}^{e}=-\text{i}\frac{1}{V}{{\left[ \frac{\partial }{\partial \dot{u}_{j}^{e}}\left( \frac{\delta {{E}_{K}}}{\delta u_{i}^{e}} \right) \right]}_{st}},$ $R_{ij}^{q}= -\text{i}\frac{1}{V}{{\left[ \frac{\partial }{\partial \dot{m}_{j}^{q}}\left( \frac{\delta {{E}_{K}}}{\delta m_{i}^{q}} \right) \right]}_{st}},$ $R_{ij}^{eq}= -\text{i}\frac{1}{V}{{\left[ \frac{\partial }{\partial \dot{m}_{j}^{q}}\left( \frac{\delta {{E}_{K}}}{\delta u_{i}^{e}} \right) \right]}_{st}}$, $K _{ij}^{e}=\sum\limits_{p,s}{{{{\tilde{k}}}_{p}}{{{\tilde{k}}}_{s}}{{\left[ \frac{\partial }{\partial u_{j,ps}^{e}}\left( \frac{d}{d{{r}_{p}}}\left( \frac{\partial \bar{\phi }}{\partial u_{i,p}^{e}} \right) \right) \right]}_{st}}},$ $K _{ij}^{eq}={{\left[ \sum\limits_{p,s}{{{{\tilde{k}}}_{p}}{{{\tilde{k}}}_{s}}}\frac{\partial }{\partial m_{j,ps}^{q}}\left( \frac{d}{d{{r}_{p}}}\frac{\partial \phi }{\partial u_{i,p}^{e}} \right)-\sum\limits_{p}{\text{i}}{{{\tilde{k}}}_{p}}\frac{\partial }{\partial m_{j,p}^{q}}\left( \frac{d}{d{{r}_{p}}}\frac{\partial \phi }{\partial u_{i,p}^{e}} \right) \right]}_{st}},$ $ K _{ij}^{q}={{\left[ \frac{\partial }{\partial m_{j}^{q}}\left( \frac{\partial \bar{\phi }}{\partial m_{i}^{q}} \right)+\sum\limits_{p}{\text{i}}{{{\tilde{k}}}_{p}}\frac{\partial }{\partial m_{j,p}^{q}}\left( \frac{\partial \phi }{\partial m_{i}^{q}} \right)-\sum\limits_{p}{\text{i}}{{{\tilde{k}}}_{p}}\frac{\partial }{\partial m_{j,p}^{q}}\right.}}$ ${{\left.\left( \frac{d}{d{{r}_{p}}}\left( \frac{\partial \phi }{\partial m_{i,p}^{q}} \right) \right)+\sum\limits_{p,s}{{{{\tilde{k}}}_{p}}{{{\tilde{k}}}_{s}}}\frac{\partial }{\partial m_{j,ps}^{q}}\left( \frac{d}{d{{r}_{p}}}\left( \frac{\partial \phi }{\partial m_{i,p}^{q}} \right) \right) \right]}_{st}}.$ Here a subscript $"st"$ means that the term is calculated at the equilibrium state ${{\mathbf{u}}^{e}}={{\left( {{\mathbf{u}}^{e}} \right)}_{st}}$ and ${{\mathbf{m}}^{q}}={{\left( {{\mathbf{m}}^{q}} \right)}_{st}}$. One should notice that the stiffness matrix $\mathbf{K}$ is completely determined by the emergent elasticity of the SkX under magnetic field\cite{20,32}. The dispersion relation for different modes ${{\omega }_{i}}={{\omega }_{i}}(\mathbf{\tilde{k}})$ can be obtained by solving eq. (\ref{4}). By defining the rescaled frequency ${{\tilde{\omega }}_{i}}=\frac{1}{\eta }{{\omega }_{i}}$, where $\eta =\left| \frac{\gamma {{m}^{3}}{{D}^{4}}}{16{{M}}{{A}^{2}}\beta } \right|$, one obtains a rescaled dispersion relation ${{\tilde{\omega }}_{i}}={{\tilde{\omega }}_{i}}(\mathbf{\tilde{k}})$ that is material-independent.
\
\\
\\
$Relation$ $between$ $the$ $emergent$ $displacement$ $and$ $the$ $Fourier$ $magnitudes$ $of$ $SkX$ $at$ $long$ $wavelength$ $limit$\\
At long wavelength limit, eq. (\ref{3}) can be expanded as
\begin{equation}
\begin{aligned}
\mathbf{m}=&\sum\limits_{\mathbf{l}}{\left[ {{\left( {{\mathbf{m}}_{{{\mathbf{q}}_{\mathbf{l}}}}} \right)}_{st}}+{{\left( {{\mathbf{m}}_{{{\mathbf{q}}_{\mathbf{l}}}}} \right)}_{v}}-\text{i}{{\left( {{\mathbf{m}}_{{{\mathbf{q}}_{\mathbf{l}}}}} \right)}_{st}}\left({\mathbf{q}}_{\mathbf{l}}\right)_{st}\cdot {{\left( {{\mathbf{u}}^{e}} \right)}_{v}} \right]} \\
&\times {{{e}^{\text{i}\left({\mathbf{q}}_{\mathbf{l}}\right)_{st}\cdot \mathbf{r}}}},
\label{15}
\end{aligned}
\end{equation}
which shows that by setting
\begin{equation}
\begin{aligned}
{{\left( {{\mathbf{m}}_{{{\mathbf{q}}_{\mathbf{l}}}}} \right)}_{v}}=\text{i}{{\left( {{\mathbf{m}}_{{{\mathbf{q}}_{\mathbf{l}}}}} \right)}_{st}}\left({\mathbf{q}}_{\mathbf{l}}\right)_{st}\cdot {{\left( {{\mathbf{u}}^{e}} \right)}_{v}}
\label{16}
\end{aligned}
\end{equation}
for all ${{\mathbf{q}}_{\mathbf{l}}}$ considered, the vibration of the whole system vanished. This means that at long wavelength limit the vibration of ${{\mathbf{u}}^{e}}$ is equivalent to a specific type of vibration of the Fourier magnitudes identified by eq. (\ref{16}). In fact, when solving the eigenvalue problem of eq. (\ref{4}), two orthogonal modes can be specified from eq. (\ref{16}), which correspond to two vanishing eigenvalues of the two matrix $\mathbf{R}$ and $\mathbf{K }$. As a result, the phonon spectrum obtained by solving eq. (\ref{4}) at long wavelength limit is equivalent to that obtained by solving the following equation
\begin{equation}
\begin{aligned}
\left( {{\mathbf{R}}^{q}}\omega -{{\mathbf{K }}^{q}} \right){{\mathbf{m}}^{q0}}=\mathbf{0}.
\label{17}
\end{aligned}
\end{equation}
The difference is that by solving eq. (\ref{17}), it is hard to distinguish between deformation of the lattice as a whole and deformation of the field pattern inside the matrix. To do this at long wavelength limit, one has to specify the two orthogonal modes from eq. (\ref{16}), which are the modes that does not cause any vibration of the system (referred to as the V-modes). The effective vibrational modes solved from eq. (\ref{4}) satisfy an orthogonal relation with the V-modes, which can be used to calculate the value of $\mathbf{u}^{e0}$ in the eigenvector of a specific mode if one solves eq. (\ref{17}) instead of eq. (\ref{4}).

\begin{acknowledgments}
The work was supported by the NSFC (National Natural Science Foundation of China) through the funds 11772360, 11472313, 11572355 and Pearl River Nova Program of Guangzhou (Grant No. 201806010134).
\end{acknowledgments}

\bibliographystyle{apsrev4-1}

\newpage
\begin{figure*}
\centering
\includegraphics[scale=0.22]{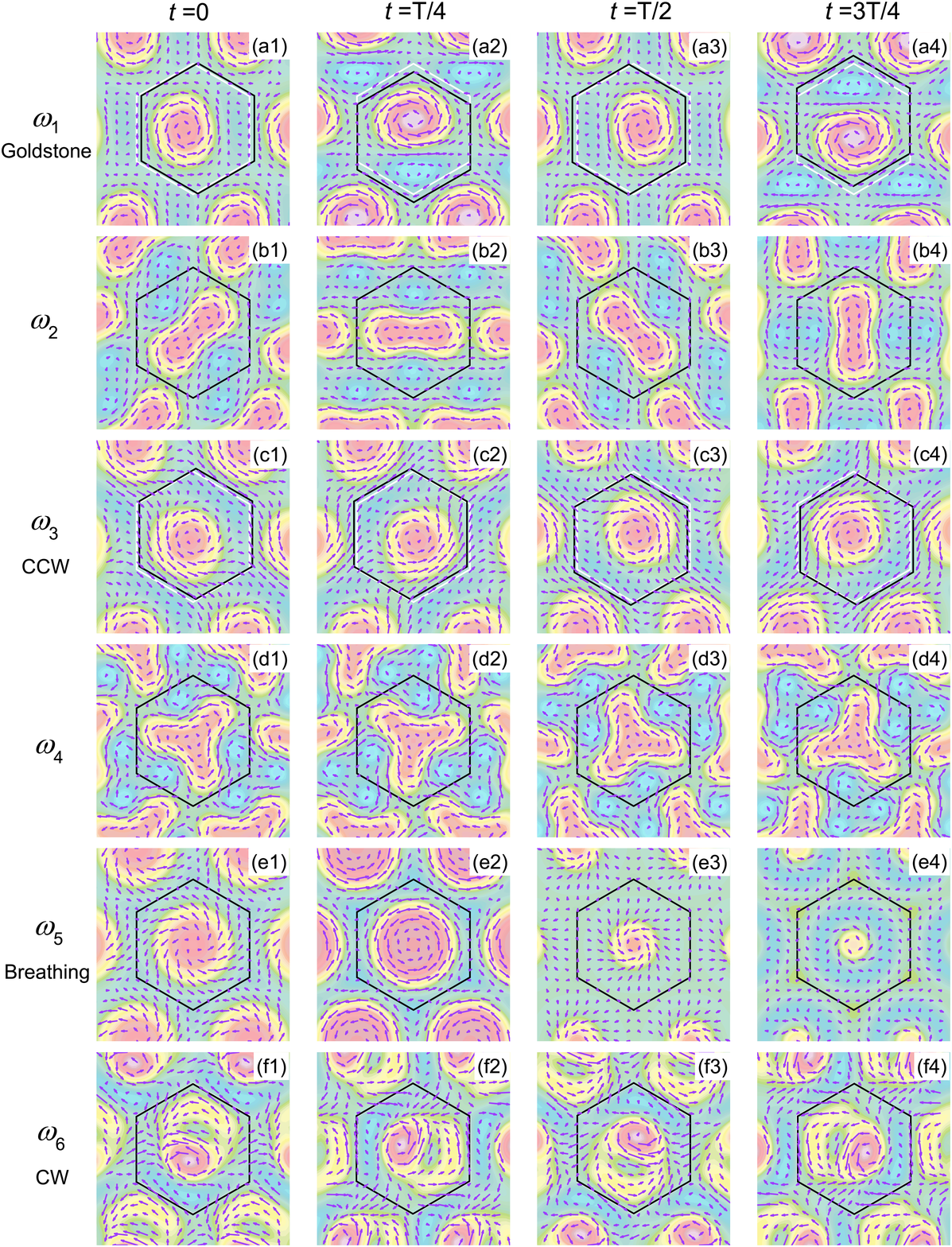}
\caption{\label{fig1}Field configurations of SkX undergoing the first 6 modes of emergent phonons calculated at $\tilde{T}=0.5,\ b=0.3$ near the $\Gamma $ point (${{\tilde{k}}_{1}}={{10}^{-\text{5}}},\ {{\tilde{k}}_{2}}=0$), different plots at four time points $t$ during a period $T$ are shown. In all figures, the vectors illustrate the distribution of the in-plane magnetization components with length proportional to their magnitude, while the colored density plot illustrates the distribution of the out-of-plane magnetization component. The black solid line plots the displaced Wigner-Seitz cell of SkX due to wave motion, while the white solid line plots the static position of the Wigner-Seitz cell. The first mode ${{\omega }_{1}}$ corresponds to the Goldstone mode; the third mode ${{\omega }_{3}}$ corresponds to the counter clockwise (CCW) mode; the fifth mode ${{\omega }_{5}}$ corresponds to the breathing mode; and the sixth mode ${{\omega }_{6}}$ corresponds to the clockwise (CW) mode.}
\end{figure*}

\begin{figure*}
\includegraphics[scale=0.35]{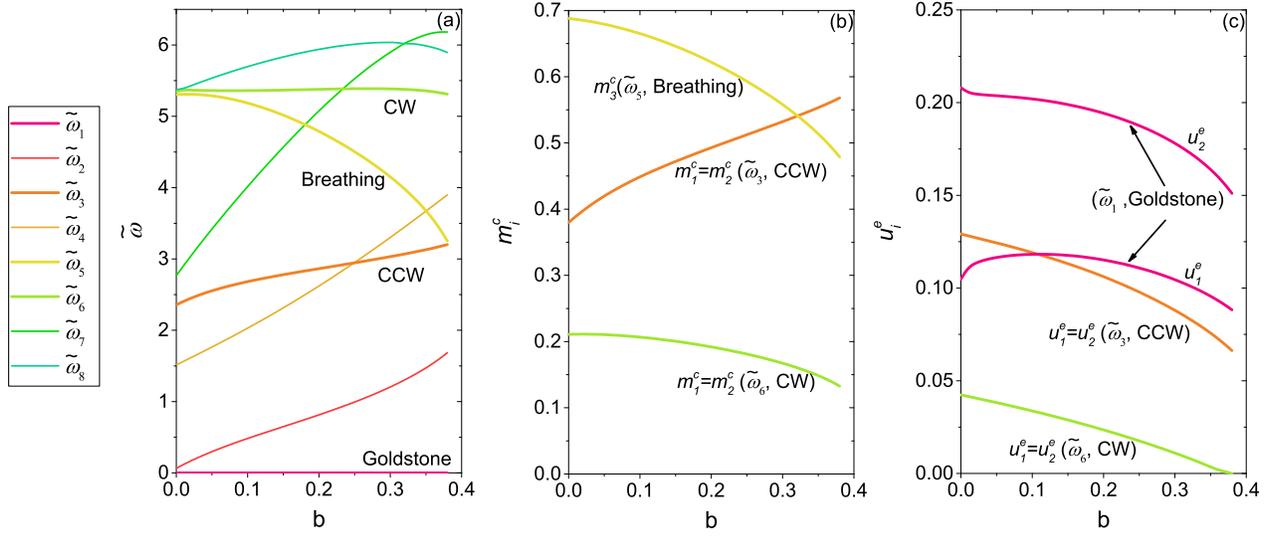}
\caption{\label{fig2}Variation of (a) the frequency, (b) the value of $m^c_i, (i=1, 2, 3)$ in the unit eigenvector, and (c) the value of $u^e_i, (i=1, 2)$ in the unit eigenvector of the first 8 modes of emergent phonons with the applied magnetic field calculated at $\tilde{T}=0.5, \tilde{k}_1=10^{-5}, \tilde{k}_2=0$.} 
\end{figure*}

\begin{figure}
\includegraphics[scale=0.34]{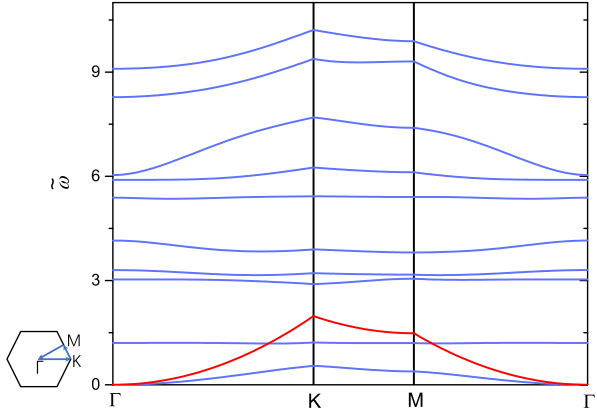}
\caption{\label{fig3}Spectrum for the first 10 modes of emergent phonons in the SkX phase of bulk helimagnets within the first Brillouin zone. The red curve plots the dispersion relation of the Goldstone mode solved by considering the vibration of $\textbf u^e$ alone.} 
\end{figure}

\begin{table}
      \caption{Specific component of the eigenvectors of the first six modes calculated at $\tilde{T}=0.5,\ b=0.3$, ${{\tilde{k}}_{1}}={{10}^{-\text{5}}},\ {{\tilde{k}}_{2}}=0$.}
      \begin{tabular}{|l|cccc|}
      \hline
      Modes & $u^{e0}_{1}$ & $u^{e0}_{2}$ & $m^{c0}_1$ ($m^{c0}_2$) & $m^{c0}_3$ \\
      \hline
      $\omega_1$(Goldstone) & 0.105 & 0.178 & 0 & 0 \\
      $\omega_2$ & 0 & 0 & 0 & 0\\
      $\omega_3$(CCW) & 0.088 & 0.088 & 0.532 & 0\\
      $\omega_4$ & 0 & 0 & 0 & 0\\
      $\omega_5$(Breathing) & 0 & 0 & 0 & 0.558\\
      $\omega_6$(CW) & 0.011 & 0.011 & 0.167 & 0\\
      $\omega_7$ & 0 & 0 & 0 & 0\\
      $\omega_8$ & 0 & 0 & 0 & 0\\
      $\omega_9$ & 0 & 0 & 0 & 0\\
      $\omega_{10}$ & 0.014 & 0.014 & 0.158 & 0\\
      $\omega_{11}$ & 0 & 0 & 0 & 0\\
      $\omega_{12}$ & 0 & 0 & 0 & 0.138\\
      $\omega_{13}$ & 0.018 & 0.018 & 0.214 & 0\\
      $\omega_{14}$ & 0.001 & 0.001 & 0.063 & 0\\
      $\omega_{15}$ & 0 & 0 & 0 & 0.281\\
      $\omega_{16}$ & 0 & 0 & 0 & 0\\
      $\omega_{17}$ & 0.008 & 0.008 & 0.132 & 0\\
      $\omega_{18}$ & 0 & 0 & 0 & 0\\
      $\omega_{19}$ & 0 & 0 & 0 & 0\\
      $\omega_{20}$ & 0 & 0 & 0 & 0\\
      \hline
      \end{tabular}
      \end{table}

\end{document}